# Exploring Fine-grained Task Parallelism on Simultaneous Multithreading Cores

Denis Los, Igor Petushkov

*Abstract*—Nowadays, latency-critical, high-performance applications are parallelized even on power-constrained client systems to improve performance. However, an important scenario of fine-grained tasking on simultaneous multithreading CPU cores in such systems has not been well researched in previous works. Hence, in this paper, we conduct performance analysis of state-of-the-art shared-memory parallel programming frameworks on simultaneous multithreading cores using real-world fine-grained application kernels. We introduce a specialized and simple software-only parallel programming framework called Relic to enable extremely fine-grained tasking on simultaneous multithreading cores. Using Relic framework, we increase performance speedups over serial implementations of benchmark kernels by 19.1% compared to LLVM OpenMP, by 31.0% compared to GNU OpenMP, by 20.2% compared to Intel OpenMP, by 33.2% compared to X-OpenMP, by 30.1% compared to oneTBB, by 23.0% compared to Taskflow, and by 21.4% compared to OpenCilk.

## I. INTRODUCTION

With an increasing amount of data to process and usage scenarios becoming more complex, modern real-world high-performance, latency-critical applications and services tend to extract more parallelism to improve performance. This trend, however, is not limited to server and high-performance computing (HPC) applications. Client devices nowadays tend to have tens of CPU cores that have enough processing power to run 3D graphics applications, rendering pipelines, and complex on-device analysis and inference engines. However, there are several differences between parallel computing in client and HPC domains.

First, power constraints on client devices are usually stricter than on HPC systems. In many cases, it is possible to simultaneously run tasks only on a few of available physical CPU cores in a client system without causing the device to overheat. CPU cores that are not utilized at the moment are put into an idle state reducing power consumption.

Second, there is a difference in the nature of fine-grained parallel tasks between client and HPC domains. In HPC domain, computations are usually divided into a big number of parallel tasks to achieve the highest performance. The spawned tasks are distributed across hundreds and thousands of CPU cores. Processing of large input data or modeling of complex multi-component systems tend to be the primary reasons for such large-scale computations in HPC domain. In client systems, a granularity of input user data is usually much smaller. Standard algorithms and operations on data structures, that are performed on a large scale in HPC systems, are applied to smaller inputs in client applications. With smaller data inputs, the same granularity of tasks could be observed by dividing the work into 2-10 independent parts rather than thousands.

Because of the stricter power constraints on client devices, an application of simultaneous multithreading (SMT) technology could help to increase a level of achieved parallelization. Simultaneous multithreading technology [1] allows to simultaneously execute instructions from multiple threads on the same physical core. Since a single thread might not fully utilize all the available resources of a CPU core due to stalls caused by events such as branch mispredictions and cache misses, running additional threads could help to improve an overall utilization of a CPU core. Most of the available commercial general-purpose processors supporting SMT implement it with 2 logical threads per 1 physical core. For instance, on x86-64 processors, Intel implements SMT technology under the name of Hyper-Threading (HT) [2], [3] supporting two logical threads per core.

The reason why SMT technology could help to parallelize applications on power-constrained client systems is that activating another physical core and scheduling a task on it consumes more power than running the task in a different logical thread on the same physical core [4]. Obviously, in most cases, using another physical core to run a parallel task is more performant [5], [6]. However, with a constrained power budget, it might not be possible leaving the utilization of SMT technology the only option to boost performance through parallelization.

Moreover, using the same physical core to run a task through SMT could help to support parallel tasks of finer granularity [7], [8]. Communication between threads is usually done through shared memory with mechanisms such as atomic operations providing synchronization. Hence, passing data through lower private levels of cache hierarchy in the same physical CPU core could reduce an overhead for scheduling parallel tasks [9]. Furthermore, there have been many works introducing and exploring hardware optimizations to reduce task scheduling and synchronization

Denis Los – Moscow Institute of Physics and Technology (9 Institutskiy per., Dolgoprudny, Moscow Region, 141700, Russian Federation)
ORCID: https://orcid.org/0009-0009-4500-8106
email: los.da@phystech.edu
Igor Petushkov – Moscow Institute of Physics and Technology (9 Institutskiy per., Dolgoprudny, Moscow Region, 141700, Russian Federation)

overheads on SMT cores [8], [10], [11].

Thus, in this paper, we focus on fine-grained tasking targeting power-constrained simultaneous multithreading CPU cores in client systems. In real-world applications, parallel computing is usually enabled through parallel programming frameworks that provide a programming interface and a runtime.

Therefore, we make the following main contributions:
1) We conduct performance analysis of fine-grained tasking in state-of-the-art shared-memory parallel programming frameworks on CPU simultaneous multithreading cores using real-world application kernels.
2) We introduce Relic: a simple specialized software-only task-parallel programming framework targeted towards enabling extremely fine-grained task parallelism on SMT cores. We show that through specialization, restrictions, and simplicity in design, we could achieve significant performance speedup compared to modern state-of-the-art parallel programming frameworks.

II. RELATED WORK

There have been many parallel programming frameworks proposed and developed throughout the years.

OpenMP [12], [13] is a standard API for shared-memory parallel programming in C/C++ and Fortran. State-of-the-art implementations of OpenMP include LLVM OpenMP [14], GNU OpenMP [15], and Intel OpenMP. X-OpenMP [16] and BOLT OpenMP [17] are more recent implementations of OpenMP introducing several optimizations to increase performance. OpenMP API allows limiting a number of worker threads and setting their thread affinity, hence, it could be used as a parallel framework to enable fine-grained tasking on SMT cores.

Many native parallel frameworks exist for C/C++ programming languages. The list includes Intel oneAPI Thread Building Blocks (oneTBB) [18], Taskflow [19], and Fastflow [20] frameworks, in which it is possible to enable fine-grained tasking on simultaneous multithreading cores.

To enable parallel programming, many solutions introduce new constructs into programming languages and require modifications to compilers. For example, OpenCilk [21] enables task-parallel programming through C/C++ language extensions. OmpSs-2 [22] introduces OpenMP-like code annotations and provides a compiler based on LLVM. Charm++ [23] adds additional functionality on top of C++ programming language. However, since large-scale client systems and applications tend to consist of many libraries and modules, using non-standard language extensions and runtimes might be challenging in real-world scenarios.

Previously, many researchers have conducted comparative performance analyses of shared-memory parallel programing frameworks, including analysis on fine-grained tasks [16], [19], [24]–[33]. Furthermore, there have been many works analyzing performance and power efficiency of parallel computing with simultaneous multithreading [5], [6], [34]–[43]. However, the scope of work on fine-grained tasking specifically on SMT cores is very limited. Most of the related studies either focus on coarse-grained or medium-grained parallelism on SMT cores or fine-grained tasks that are spawned in large numbers from parallelizing heavy workloads, and thus are also scheduled on SMT cores. Moreover, to the best of our knowledge, there have not been works that conduct performance analysis of multiple state-of-the-art parallel programming frameworks on SMT cores using fine-grained tasks that come from workloads with small input datasets.

In [34], [40], [41], and [43], using NAS Parallel Benchmarks [44], performance evaluations are done for parallel computing on x86 CPU cores with hyper-threading. However, the analysis is limited to OpenMP and coarse-grained application kernels. In [39], performance analysis of tasking on SMT cores is also conducted based on NAS Parallel Benchmarks with large inputs, however, explicit threading is used instead of OpenMP. Performance and power efficiency of x86-64 Intel processors with hyper-threading is explored in [38] using SPEC OMP [45] and SPEC CPU2006 [46] benchmarks with standard reference inputs. Acceleration of applications from different domains using SMT technology is studied in [5], [6], [35]–[37], and [42], but the focus is also on the coarse-grained parallelism.

Modern state-of-the-art task-parallel frameworks such as OpenMP, Intel oneAPI Thread Building Blocks, OpenCilk, and Taskflow do not provide specialized constructs or interfaces that could be used to reduce task handling overheads and enable extremely fine-grained tasking specifically on simultaneous multithreading cores. However, mechanisms that could be utilized to reduce task handling overheads on SMT cores have been studied in previous works.

SMT technology does not only allow to speedup applications through parallelization but also to use other logical threads of a physical core as helper threads [47]. Helper threads could be used to speculatively prefetch data for the main thread or precompute conditions for hard-to-predict branches [39], [47]. A lot of the techniques reducing task handling and synchronization overheads have been discussed primarily in the context of the helper thread scenario. The techniques that enable extremely fine-grained tasking can be either hardware-accelerated [8], [10], [11], [48]–[50] or software-only [16], [39], [51], [52].

Some of the software-only techniques are not limited to the usage on SMT cores and have already been utilized in state-of-the-art task-parallel frameworks, such as X-OpenMP, to reduce task handling overheads [16]. For example, in [16] and [39], different implementations of synchronization primitives and worker thread suspension mechanisms are studied to enable fine-grained tasking.

Furthermore, several works introduce novel hardware-accelerated general parallel computing models and microarchitectures to enable extremely fine-grained tasking and speculative parallelism beyond simultaneous multithreading cores [53]–[55].

Despite the large scope of work on hardware-accelerated parallel computing models and task handling optimizations, almost none of them are supported in current commercially available processors with simultaneous multithreading [39]. To enable fine-grained tasking in applications, aforementioned software-only parallel programming frameworks are used in both client devices and HPC systems. Even though, several software-only mechanisms that could help to enable extremely fine-grained tasking specifically on SMT cores had been previously researched, to the best of knowledge, there were not any studies conducted highlighting achievable performance speedups over popular existing software-only task-parallel frameworks. Hence, in this paper, we introduce a specialized task-parallel framework utilizing software-only techniques to enable extremely fine-grained parallelism on SMT cores and demonstrate significant performance speedups over modern state-of-the-art general parallel programming frameworks.

III. METHODOLOGY

For all performance evaluations, we use a computer with Intel Core i7-8700 @ 3.20 GHz x86-64 processor featuring 6 physical CPU cores with 2 logical threads per each core. The system is Ubuntu 22.04 with Linux 5.15 kernel and glibc 2.35.

We evaluate four state-of-the-art OpenMP implementations: LLVM OpenMP from LLVM 18.1.2, GNU OpenMP from GCC 13.2, Intel OpenMP from Intel oneAPI Base Toolkit 2024.0, and X-OpenMP. We ported the original X-OpenMP implementation from LLVM 11 to 18.1.2. We also evaluate Intel oneAPI Thread Building Blocks (oneTBB) from the 2021.11 release, OpenCilk from the 2.1 release, and Taskflow v3.7.0 parallel programming system.

For OpenMP implementations, we use #pragma task and #pragma taskwait directives to submit a task and wait for it to finish. We use the task_group class and its methods with oneTBB. In Taskflow parallel programing system, we rely on asynchronous tasking. For OpenCilk, we use cilk_spawn and cilk_sync standard calls.

Benchmarks and investigated parallel runtimes are compiled with LLVM 18.1.2 Clang and -O3 optimization options. However, we use GCC 13.2 compiler to evaluate GNU OpenMP implementation since it is not compatible with LLVM. Moreover, OpenCilk 2.1 is based on LLVM 16.0.6. We use LLVM's libc++ standard library implementation from LLVM 18.1.2 as the default for all cases.

In order to conduct performance analysis on SMT core, we limit a number of worker threads to 2 for each runtime and bind them to the same physical CPU core.

IV. BENCHMARKS

We focus on evaluating performance of parallel runtimes and frameworks enabling fine-grained tasking on SMT cores. Many previous studies have showed that performance gains from SMT technology greatly depend on applications [38], [39], [56]. However, since, in general, parallel memory-intensive tasks with complex memory access patterns are more likely to benefit from running in logical threads of a SMT core [39], we primarily use real-world fine-grained memory-intensive application kernels for performance evaluation.

*A. Graph algorithms*

Graph algorithms are standard building blocks used in many client applications. We choose betweenness centrality (BC), breadth-first search (BFS), connected components (CC), page ranking (PR), single-source shortest paths (SSSP), and triangle counting (TC) graph kernels for performance evaluation. We take single-threaded high-performance implementations of these graph kernels from GAP Benchmark Suite [57]. For the connected components graph kernel, we use the implementation based on Shiloach-Vishkin algorithm [58], since it shows better performance on fine-grained input graphs.

In order to evaluate performance on SMT cores, we run two instances of the same graph kernel in parallel, binding them to different logical threads of a physical core. Each instance operates on the same input graph. Basically, we generate two identical graphs and pass them to graph kernel instances. The tasks are scheduled using a parallel runtime under investigation. In the serial mode, we run two instances of a graph kernel in a single thread.

We use a generated Kronecker graph with 32 nodes and 157 undirected edges for a degree of 4 as an input for all graph kernels. With this generated graph used as an input, a single task instance takes 1.1 microseconds for the BC graph kernel, 0.5 microseconds for the BFS kernel, and 0.4 microseconds for the CC kernel. For PR, SSSP, and TC graph kernels, task instances take 4.3 microseconds, 6.4 microseconds, and 1.3 microseconds to compute, respectively. Since we use fine-grained tasks, we repeat the experiments for $10^5$ iterations and average the results to improve stability of performance measurements.

*B. JSON parsing*

JavaScript Object Notation (JSON) is a standard format that is frequently used to transmit data between client applications and web servers. Hence, parsing of received JSON files in parallel could improve performance of client applications.

To conduct performance evaluation for JSON parsing scenario, we use RapidJSON library [59]. RapidJSON is a fast C++ library for parsing and generating JSON files. As an input file, we use a small sample JSON file that describes a widget and is available from [60]. We run two tasks both parsing this JSON file loaded into a memory buffer. Each task has its own copy of the memory buffer with the loaded file content. With the selected JSON file as an input, a single JSON parsing task takes 1.1 microseconds to complete in our testing environment. We bind each task to different logical threads on the same SMT physical core. However, in the serial mode, we run these two JSON parsing tasks in the same thread. We run all the experiments for $10^5$ iterations and average the results.

V. PERFORMANCE ANALYSIS OF STATE-OF-THE-ART PROGRAMMING FRAMEWORKS

We conduct performance analysis of the state-of-the-art task-parallel frameworks using aforementioned real-world fine-grained application kernels consisting of graph algorithms and JSON parsing with small inputs. In Fig. 1, performance speedups over serial implementations are presented for each of the investigated parallel frameworks.

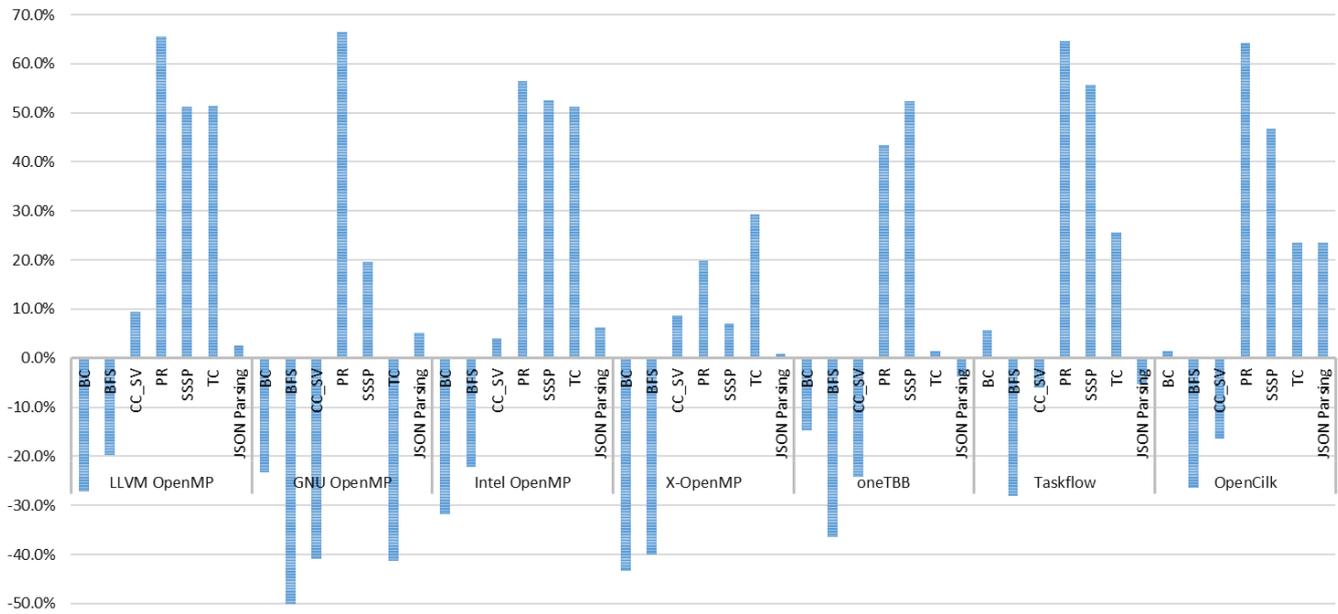

Figure 1. Performance speedups over serial implementations of application kernels with different state-of-the-art parallel programming frameworks

LLVM OpenMP shows the best performance speedup not only among different OpenMP implementations, but among all of the investigated frameworks. Using the geometric mean to average the results across benchmarks and taking into account performance degradations on individual application kernels, LLVM OpenMP shows 13.9% performance speedup over serial implementations, while Intel OpenMP, Taskflow, and OpenCilk show 11.3%, 11.8%, and 12.6% speedups, respectively. However, averaging across all the application kernels, X-OpenMP, GNU OpenMP, and oneTBB frameworks show 6.7%, 17.7%, and 1.9% performance degradations, respectively.

All the frameworks achieve performance speedups on the PR and SSSP benchmark kernels, the ones with the highest granularities. Moreover, on the triangle counting graph kernel, GNU OpenMP is the only one that results in performance degradation. However, only Taskflow and OpenCilk are able to achieve performance speedups on the BC graph kernel, while none of the parallel frameworks could successfully parallelize the benchmark using breath-first search algorithm. Furthermore, among the investigated parallel frameworks, only LLVM-based OpenMP implementations show increase in performance on the connected components graph kernel. JSON parsing benchmark is successfully parallelized with all the investigated OpenMP implementations and OpenCilk parallel framework.

## VI. RELIC: A SPECIALIZED FRAMEWORK FOR FINE-GRAINED TASKING ON SMT CORES

We introduce Relic, a specialized framework for C and C++ programming languages to enable extremely fine-grained task parallelism on SMT cores.

### A. Task scheduling

State-of-the-art shared-memory parallel programming frameworks such as oneTBB, OpenCilk, and OpenMP implementations tend to use advanced work-stealing algorithms to efficiently distribute tasks across tens and hundreds of CPU cores. However, a parallel programming framework that is specialized to the usage on a SMT core needs to distribute tasks only among available logical threads. In most cases, there are only two logical threads on a SMT core. Hence, in this paper, we consider only the case with 2 running logical threads.

To remove the necessity to implement any complex scheduling strategies and reduce a task scheduling overhead, we assign special roles to each of the two threads. One of the threads is made the *main* thread, while the other – the *assistant* thread. The main thread is one of the primary application threads created by an application itself or by a runtime of a general-purpose parallel programming framework. The assistant thread is created and managed by Relic task-parallel framework. The main thread is a producer and the assistant thread is a consumer, meaning that only the main thread can submit tasks in Relic, while the assistant thread is the only one allowed to run them. The assistant thread cannot submit tasks, hence, creating tasks recursively is not supported in Relic.

Thus, in Relic, we implement the single-producer single-consumer pattern. This pattern and its utilization to reduce task scheduling overheads in parallel runtimes have been well studied in previous works, for example, in [16] and [20]. To submit tasks, we use a single-producer single-consumer (SPSC) lock-free queue, a standard mechanism to utilize in such scenario. There have been many SPSC queue implementations proposed throughout the years [61]–[64]. In this paper, we use the SPSC queue implementation available in Boost C++ libraries [65]. We set a capacity of the queue to 128 entries.

147

In Relic framework, a task can be submitted by calling the submit() function in the main thread passing pointers to a task routine and its arguments. To wait for the completion of all currently submitted tasks, the wait() function should be used in the main thread.

Relic framework could be used together with a general-purpose parallel programming framework. Coarse-grained or medium-grained tasks could be submitted to the main thread through a general-purpose parallel framework, while further extremely fine-grained parallelization of these tasks within the same physical CPU core could be enabled with Relic framework.

### B. Waiting mechanism and OS thread scheduling

If there are no tasks in the SPSC queue, the assistant thread will wait for the main thread to submit the work. Moreover, the main thread can wait for the assistant thread to finish execution of the submitted tasks. These waiting mechanisms can be implemented in many different ways. However, in general, there are two distinct approaches: busy-waiting and thread suspension. With busy-waiting or spinning, a waiting thread checks for a condition to become true in a loop consuming CPU cycles. Alternatively, a thread can suspend the execution and release a CPU core using one the mechanisms provided by the operating system. There is also a hybrid approach, in which a thread spins for a short interval and then suspends its execution.

Both spinning and thread suspension mechanisms have been well studied. It is known that spinning tends to show better performance for short waiting intervals in lightly contended environments [66]. In our case, we have only two communicating threads running on the same physical core and we focus on supporting extremely fine-grained tasks. Therefore, in Relic, we use busy-waiting in the main and assistant threads. For x86-64 machines, we use the pause instruction to make spinning more efficient. In Fig. 2, pseudocode for the main loop of the assistant thread is shown.

```
while true do
    while SPSCTaskQueueReadAvailable() is false do
        Pause()
    end
    TaskRoutine, Args ← SPSCTaskQueueFront();
    TaskRoutine(Args);
    SPSCTaskQueuePop();
end
```

Figure 2. Pseudocode for the main loop of the assistant thread

However, in real-world high-performance, latency-critical client applications, only a part of the system could usually be parallelized. It means that the assistant thread might end up waiting in the busy loop for longer durations of time. Hence, it could make spinning extremely inefficient and degrade performance of the whole client application. One of the possible solutions is to use the hybrid approach. However, with fine-grained tasks, the overhead from awakening the assistant thread might outweigh performance benefits from the parallelization.

In [66], an optimization to mitigate thread awakening overheads is suggested, however, adopting this technology directly to our scenario with only one waiting thread is challenging. In Relic, we use a different approach. Since detailed profiling is usually conducted for critical applications, we leave it to application developers to provide explicit wake-up and sleep hints to the runtime. We provide wake_up_hint() and sleep_hint() functions for developers to call some time before and after parallelizable code sections in applications in order to wake up and suspend the assistant thread, respectively. This way, we enable a fine-grained control over the assistant thread in Relic.

With Relic, we enable fine-grained tasking on SMT cores and we expect the main and assistant threads to run on the same physical core. Relic framework will work correctly, if the threads are scheduled to different physical cores, however, it is not intended or optimized for such scenario. We do not implement the CPU pinning algorithms in Relic and expect users of the framework to set the CPU affinities for both the main and assistant threads. Either simple static thread binding schemes or complex dynamic scheduling strategies could be implemented by application developers to support all scenarios present in the target applications.

## VII. RESULTS

In Fig. 3, performance speedups over serial implementations of investigated application kernels are presented for Relic parallel programming framework.

All of the investigated fine-grained benchmarks are successfully parallelized with Relic without performance degradations. Even the benchmark utilizing BFS algorithm is accelerated by 5.6% using Relic parallel framework. On average, Relic parallel programming framework shows 42.1% performance speedup over serial implementations.

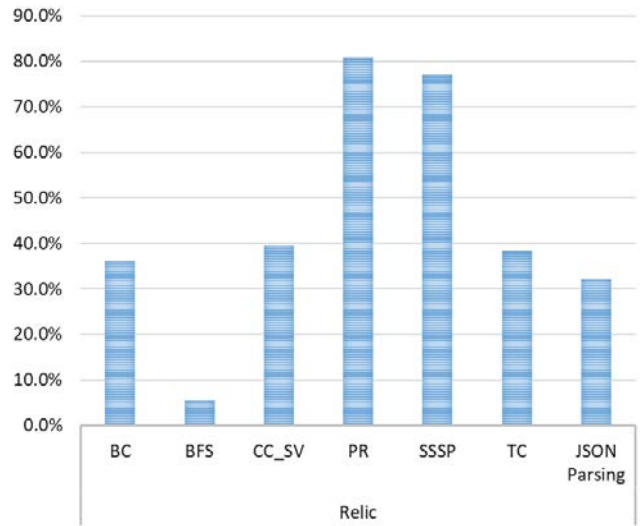

Figure 3. Performance speedups over serial implementations of application kernels with Relic framework

The best achievable performance speedups over serial implementations with previously evaluated parallel programming frameworks are 5.7% for the BC benchmark kernel with Taskflow, 9.4% for the CC benchmark kernel with LLVM OpenMP, 66.5% for the PR benchmark kernel with GNU OpenMP, 55.7% for the SSSP benchmark kernel with Taskflow, 51.4% for the TC benchmark kernel with LLVM OpenMP, and 23.5% for the JSON parsing benchmark with OpenCilk. Relic parallel programming framework increases achievable performance speedups by 30.4%, 30.1%, 14.3%, 21.3%, and 8.6% for the BC, CC, PR, SSSP, and JSON parsing benchmarks, respectively. Only for the benchmark using the triangle counting algorithm, Relic framework shows lower performance speedup compared to the best one that is achieved using LLVM OpenMP.

As we have mentioned before, extensive profiling and performance analysis are usually conducted for real-world latency-critical applications. Therefore, performance degradations from the introduced parallelization for specific code sections would be discovered and addressed by reverting back to serial implementations. Thus, in Fig. 4, for the investigated task-parallel frameworks, performance speedups over serial implementations are shown without negative outliers, using the geometric mean to average the results. In case of the performance degradation on a specific benchmark kernel, a result for the baseline serial implementation is used. Hence, Relic parallel programming framework increases the performance benefits from the parallelization by 19.1% compared to LLVM OpenMP, by 31.0% compared to GNU OpenMP, by 20.2% compared to Intel OpenMP, by 33.2% compared to X-OpenMP, by 30.1% compared to oneTBB, by 23.0% compared to Taskflow, and by 21.4% compared to OpenCilk.

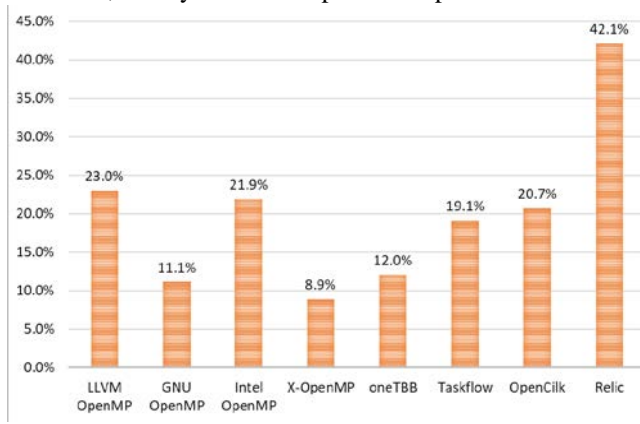

Figure 4. Average performance speedups across application kernels without negative outliers with different parallel programming frameworks

## VIII. CONCLUSION

We conduct performance analysis of seven state-of-the-art shared-memory parallel programming frameworks on a simultaneous multithreading CPU core using real-world fine-grained application kernels consisting of graph algorithms and JSON parsing. We show performance degradations on several investigated fine-grained tasks with the existing task-parallel frameworks.

We introduce Relic, a simple specialized parallel programming framework enabling extremely fine-grained task parallelism on simultaneous multithreading cores. With Relic framework, we demonstrate significant performance improvements compared to the existing general-purpose parallel frameworks.

REFERENCES

[1] D. M. Tullsen, S. J. Eggers, and H. M. Levy, "Simultaneous multithreading: maximizing on-chip parallelism," in *Proc. 22nd Annual International Symposium on Computer Architecture*, Santa Margherita Ligure, Italy, 1995, pp. 392-403.
[2] D. T. Marr *et al*., "Hyper-Threading technology architecture and microarchitecture," *Intel Technology Journal*, vol. 6, no. 1, pp. 4-15, 2002.
[3] D. Koufaty and D. T. Marr, "Hyperthreading technology in the netburst microarchitecture," *IEEE Micro*, vol. 23, no. 2, pp. 56-65, March-April 2003, DOI: 10.1109/MM.2003.1196115.
[4] Y. Zhai, X. Zhang, S. Eranian, L. Tang, and J. Mars, "HaPPy: hyperthread-aware power profiling dynamically," in *Proc. of the 2014 USENIX Conference on USENIX Annual Technical Conference,* Philadelphia, PA, USA, 2014, pp. 211-218.
[5] T. Leng, R. Ali, J. Hsieh, V. Mashayekhi, and R. Rooholamini, "An empirical study of hyper-threading in high performance computing clusters," *Linux HPC Revolution*, Article ID 45, 2002.
[6] L. Pons *et al*., "Effect of hyper-threading in latency-critical multithreaded cloud applications and utilization analysis of the major system resources," *Future Gener. Comput. Syst.*, vol. 131, pp. 194-208, June 2022.
[7] N. Tuck and D. M. Tullsen, "Initial observations of the simultaneous multithreading Pentium 4 processor," in *2003 12th International Conference on Parallel Architectures and Compilation Techniques*, New Orleans, LA, USA, 2003, pp. 26-34, DOI: 10.1109/PACT.2003.1237999.
[8] D. M. Tullsen, J. L. Lo, S. J. Eggers, and H. M. Levy, "Supporting fine-grained synchronization on a simultaneous multithreading processor," in *Proc. Fifth International Symposium on High-Performance Computer Architecture*, Orlando, FL, USA, 1999, pp. 54-58, DOI: 10.1109/HPCA.1999.744326.
[9] X. Qian, B. Sahelices, and J. Torrellas, "BulkSMT: Designing SMT processors for atomic-block execution," in *IEEE International Symposium on High-Performance Comp Architecture*, New Orleans, LA, USA, 2012, pp. 1-12, DOI: 10.1109/HPCA.2012.6168952.
[10] N. Anastopoulos and N. Koziris, "Facilitating efficient synchronization of asymmetric threads on hyper-threaded processors," in *2008 IEEE International Symposium on Parallel and Distributed Processing*, Miami, FL, USA, 2008, pp. 1-8, DOI: 10.1109/IPDPS.2008.4536358.
[11] J. L. Kihm and D. A. Connors, "Implementation of fine-grained cache monitoring for improved SMT scheduling," in *IEEE International Conference on Computer Design: VLSI in Computers and Processors, 2004. ICCD 2004. Proceedings.,* San Jose, CA, USA, 2004, pp. 326-331, DOI: 10.1109/ICCD.2004.1347941.
[12] L. Dagum and R. Menon, "OpenMP: an industry standard API for shared-memory programming," *IEEE Computational Science and Engineering*, vol. 5, no. 1, pp. 46-55, Jan.-March 1998, DOI: 10.1109/99.660313.
[13] E. Ayguade *et al*., "The design of OpenMP tasks," *IEEE Transactions on Parallel and Distributed Systems*, vol. 20, no. 3, pp. 404-418, March 2009, DOI: 10.1109/TPDS.2008.105.
[14] O. A. R. Board, OpenMP, Support for the OpenMP language, 2024. [Online]. Available: https://openmp.llvm.org.
[15] G.team, Gomp: An openmp implementation for gcc, 2024. [Online]. Available: https://gcc.gnu.org/projects/gomp.
[16] P. Nookala, K. Chard, I. Raicu, "X-OpenMP — eXtreme fine-grained tasking using lock-less work stealing," *Future Generation Computer Systems*, vol. 159, pp. 444-458, 2024, DOI: 10.1016/j.future.2024.05.019.
[17] S. Iwasaki, A. Amer, K. Taura, S. Seo and P. Balaji, "BOLT: optimizing OpenMP parallel regions with user-level threads," in *2019 28th International Conference on Parallel Architectures and Compilation Techniques (PACT)*, Seattle, WA, USA, 2019, pp. 29-42, DOI: 10.1109/PACT.2019.00011.
[18] A. Kukanov, M. J. Voss, "The foundations for scalable multi-core software in Intel Threading Building Blocks.," *Intel Technology Journal*, vol. 11, no. 4, p. 309, 2007.
[19] T. -W. Huang, Y. Lin, C. -X. Lin, G. Guo, and M. D. F. Wong, "Cpp-Taskflow: a general-purpose parallel task programming system at scale," *IEEE Transactions on Computer-Aided Design of Integrated Circuits and Systems*, vol. 40, no. 8, pp. 1687-1700, Aug. 2021, DOI: 10.1109/TCAD.2020.3025075.
[20] M. Aldinucci, M. Danelutto, P. Kilpatrick, M. Torquati, "Fastflow: high-Level and efficient streaming on multicore," *Programming*

*Multicore and Many-Core Computing Systems,* Wiley-Blackwell, 2017, pp. 261-280, DOI: 10.1002/9781119332015.ch13.

[21] T. B. Schardl and I-T. A. Lee, "OpenCilk: a modular and extensible software infrastructure for fast task-parallel code", in *Proc. of the 28th ACM SIGPLAN Annual Symposium on Principles and Practice of Parallel Programming*, Montreal, QC, Canada, 2024, pp. 189-203, DOI: 10.1145/3572848.3577509.

[22] Barcelona Supercomputing Center, OmpSs-2 Specification, 2024. [Online]. Available: https://pm.bsc.es/ftp/ompss-2/doc/spec.

[23] L.V. Kale and S. Krishnan, "CHARM++: a portable concurrent object oriented system based on C++", in *Proc. of the Eighth Annual Conference on Object-Oriented Programming Systems, Languages, and Applications,* Washington, D.C., USA, 1993, pp. 91-108, DOI: 10.1145/165854.165874.

[24] A. Podobas, M. Brorsson, and K.-F. Faxén, "A comparison of some recent task-based parallel programming models," in *3rd workshop on programmability issues for multi-core computers*, Pisa, Italy, 2010.

[25] G.W. Price, D. K. Lowenthal, "A comparative analysis of fine-grain threads packages," *Journal of Parallel and Distributed Computing*, vol. 63, no. 11, pp. 1050-1063, 2003.

[26] K. Wheeler, D. Stark, and R. Murphy, "A comparative critical analysis of modern task-parallel runtimes," Sandia National Laboratories, Albuquerque, New Mexico, USA, SAND2012-10594, Dec. 2012.

[27] A. Podobas, M. Brorsson, and K.-F. Faxen, "A comparative performance study of common and popular task-centric programming frameworks," *Concurr. Comput.: Pract. Exper.*, vol. 27, no. 1, pp. 1-28, Jan. 2015, DOI: 10.1002/cpe.3186.

[28] G. Zeng, "Performance analysis of parallel programming models for C++," *J. Phys.: Conf. Ser.,* vol. 2646, 2023, DOI: 10.1088/1742-6596/2646/1/012027.

[29] E. Ajkunic, H. Fatkic, E. Omerovic, K. Talic, and N. Nosovic, "A comparison of five parallel programming models for C++," in *2012 Proceedings of the 35th International Convention MIPRO*, Opatija, Croatia, 2012, pp. 1780-1784.

[30] A. Leist, A. Gilman, "A comparative analysis of parallel programming models for C++," in *Proc. of The Ninth International Multi-Conference on Computing in the Global Information Technology*, Seville, Spain, 2014, pp. 121-127.

[31] C. D. Krieger, M. M. Strout, J. Roelofs, and A. Bajwa, "Executing optimized irregular applications using task graphs within existing parallel models," in *2012 SC Companion: High Performance Computing, Networking Storage and Analysis*, Salt Lake City, UT, USA, 2012, pp. 261-268, DOI: 10.1109/SC.Companion.2012.43.

[32] L.M. Sanchez, J. Fernandez, R. Sotomayor, S. Escolar, J.D. Garcia, "A comparative study and evaluation of parallel programming models for shared-memory parallel architectures," *New Gener. Comput.*, vol. 31, pp. 139–161, 2013, DOI: 10.1007/s00354-013-0301-5.

[33] S. Salehian, Jiawen Liu, and Yonghong Yan, "Comparison of Threading Programming Models," in *2017 IEEE International Parallel and Distributed Processing Symposium Workshops (IPDPSW)*, Lake Buena Vista, FL, USA, 2017, pp. 766-774, DOI: 10.1109/IPDPSW.2017.141.

[34] W. Heirman, T. E. Carlson, K. Van Craeynest, I. Hur, A. Jaleel, L. Eeckhout, "Automatic SMT threading for OpenMP applications on the Intel Xeon Phi co-processor," in *Proc. of the 4th International Workshop on Runtime and Operating Systems for Supercomputers*, Munich, Germany, 2014, Article 7, DOI: 10.1145/2612262.2612268

[35] X. Tian, Y.-K. Chen, M. Girkar, S. Ge, R. Lienhart and S. Shah, "Exploring the use of Hyper-Threading technology for multimedia applications with Intel OpenMP compiler," in *Proc. International Parallel and Distributed Processing Symposium*, Nice, France, 2003, DOI: 10.1109/IPDPS.2003.1213118.

[36] Y.-K. Chen, M. Holliman, E. Debes, S. Zheltov, A. Knyazev, S. Bratanov, R. Belenov, and I. Santos, "Media applications on Hyper-Threading technology," *Intel Technology Journal*, vol. 6, no. 1, pp. 47-57, 2002.

[37] Y.-K. Chen, M. Holliman, and E. Debes, "Video applications on hyper-threading technology," in *Proc. IEEE International Conference on Multimedia and Expo*, Lausanne, Switzerland, 2002, pp. 193-196, vol. 2, DOI: 10.1109/ICME.2002.1035546.

[38] R. Schöne, D. Hackenberg, and D. Molka, "Simultaneous multithreading on x86_64 systems: an energy efficiency evaluation," in *Proc. of the 4th Workshop on Power-Aware Computing and Systems*, Cascais, Portugal, 2011, Article 10, DOI: 10.1145/2039252.2039262.

[39] E. Athanasaki, N. Anastopoulos, K. Kourtis, N. Koziris, "Exploring the performance limits of simultaneous multithreading for memory intensive applications," *The Journal of Supercomputing*, vol. 44, pp. 64-97, 2008, DOI: 10.1007/s11227-007-0149-x.

[40] E. Athanasaki, N. Anastopoulos, K. Kourtis, N. Koziris, "Exploring the capacity of a modern SMT architecture to deliver high scientific application performance," in *Proc. of the 2006 International Conference on High Performance Computing and Communications*, Munich, Germany, 2006, pp. 180-189, DOI: 10.1007/11847366_19.

[41] R. E. Grant and A. Afsahi, "A Comprehensive Analysis of OpenMP Applications on Dual-Core Intel Xeon SMPs," in *2007 IEEE International Parallel and Distributed Processing Symposium*, Long Beach, CA, USA, 2007, pp. 1-8, DOI: 10.1109/IPDPS.2007.370682.

[42] S. Ivanikovas and G. Dzemyda, "Evaluation of the hyper-threading technology for heat conduction-type problems," *Mathematical Modeling and Analysis*, vol. 12, no. 4, pp. 459-468, Dec. 2007.

[43] M. Curtis-Maury, X. Ding, C.D. Antonopoulos, D.S. Nikolopoulos, "An evaluation of OpenMP on current and emerging multithreaded/multicore processors," in *Proc. of the First International Workshop on OpenMP Shared Memory Parallel Programming*, Eugene, OR, USA, 2005, pp. 133-144.

[44] H. Jin, M. Frumkin, and J. Yan, "The OpenMP implementation of NAS Parallel Benchmarks and its performance," NASA Ames Research Center, Technical Report, Oct. 1999.

[45] V. Aslot, M. J. Domeika, R. Eigenmann, G. Gaertner, W. B. Jones, and B. Parady, "Specomp: A new benchmark suite for measuring parallel computer performance," in *Proc. of the International Workshop on OpenMP Applications and Tools: OpenMP Shared Memory Parallel Programming*, London, UK, 2001, pp. 1-10

[46] J. L. Henning, "SPEC CPU2006 benchmark descriptions," *SIGARCH Comput. Archit. News*, vol. 34, no. 4, pp. 1-17, Sep. 2006, DOI: 10.1145/1186736.1186737.

[47] J. D. Collins *et al.*, "Speculative precomputation: long-range prefetching of delinquent loads," in *Proc. 28th Annual International Symposium on Computer Architecture*, Gothenburg, Sweden, 2001, pp. 14-25, DOI: 10.1109/ISCA.2001.937427.

[48] A. Gontmakher, A. Mendelson, A. Schuster and G. Shklover, "Speculative synchronization and thread management for fine granularity threads," in *The Twelfth International Symposium on High-Performance Computer Architecture, 2006.*, Austin, TX, USA, 2006, pp. 278-287, DOI: 10.1109/HPCA.2006.1598136.

[49] J. Redstone, S. Eggers and H. Levy, "Mini-threads: increasing TLP on small-scale SMT processors," in *The Ninth International Symposium on High-Performance Computer Architecture, 2003. HPCA-9 2003. Proceedings.*, Anaheim, CA, USA, 2003, pp. 19-30, DOI: 10.1109/HPCA.2003.1183521.

[50] M. Abeydeera, S. Subramanian, M. C. Jeffrey, J. Emer and D. Sanchez, "SAM: Optimizing Multithreaded Cores for Speculative Parallelism," in *2017 26th International Conference on Parallel Architectures and Compilation Techniques (PACT)*, Portland, OR, USA, 2017, pp. 64-78, DOI: 10.1109/PACT.2017.37.

[51] K.-F. Faxén, "Wool-a work stealing library," *SIGARCH Comput. Archit. News*, vol. 36, no. 5, pp. 93-100, Dec. 2008, DOI: 10.1145/1556444.1556457.

[52] R. Rangan *et al.*, "Speculative Decoupled Software Pipelining," in *Proc. of the 16th International Conference on Parallel Architecture and Compilation Techniques*, Brasov, Romania, 2007, pp. 49-59.

[53] M. C. Jeffrey, S. Subramanian, C. Yan, J. Emer and D. Sanchez, "A scalable architecture for ordered parallelism," in *Proc. of the 48th Annual IEEE/ACM International Symposium on Microarchitecture*, Waikiki, HI, USA, 2015, pp. 228-241, DOI: 10.1145/2830772.2830777.

[54] M. C. Jeffrey, S. Subramanian, C. Yan, J. Emer and D. Sanchez, "Unlocking Ordered Parallelism with the Swarm Architecture," *IEEE Micro*, vol. 36, no. 3, pp. 105-117, May-June 2016, DOI: 10.1109/MM.2016.12.

[55] S. Kumar, C. J. Hughes, and A. Nguyen, "Carbon: architectural support for fine-grained parallelism on chip multiprocessors," in *Proc. of the 34th Annual International Symposium on Computer Architecture*, San Diego, California, USA, 2007, pp. 162-173, DOI: 10.1145/1250662.1250683.

[56] S. Saini, A. Naraikin, R. Biswas, D. Barkai and T. Sandstrom, "Early performance evaluation of a "Nehalem" cluster using scientific and engineering applications," in *Proc. of the Conference on High Performance Computing Networking, Storage and Analysis*, Portland, OR, USA, 2009, pp. 1-12, DOI: 10.1145/1654059.1654084.

[57] S. Beamer, K. Asanović, D. Patterson, "The GAP benchmark suite," arXiv:1508.03619 [cs.DC], 2015.
[58] Y. Shiloach and U. Vishkin, "An o(logn) parallel connectivity algorithm," *Journal of Algorithms*, vol. 3, no. 1, pp. 57-67, 1982.
[59] RapidJSON library, 2024. [Online]. Available: https://rapidjson.org.
[60] JSON Example, 2024. [Online]. Available: https://json.org/example.html.
[61] L. Lamport, "Specifying Concurrent Program Modules," *ACM Trans. Program. Lang. Syst.*, vol. 5, no. 2, pp. 190-222, 1983, DOI: 10.1145/69624.357207.
[62] P. P. C. Lee, T. Bu, and G. Chandranmenon, "A lock-free, cache-efficient multi-core synchronization mechanism for line-rate network traffic monitoring," in *2010 IEEE International Symposium on Parallel & Distributed Processing (IPDPS)*, Atlanta, GA, USA, 2010, pp. 1-12, DOI: 10.1109/IPDPS.2010.5470368.
[63] J. Giacomoni, T. Moseley, and M. Vachharajani, "FastForward for efficient pipeline parallelism: a cache-optimized concurrent lock-free queue," in *Proc. of the 13th ACM SIGPLAN Symposium on Principles and Practice of Parallel Programming*, Salt Lake City, UT, USA, 2008, pp. 43-52, DOI: 10.1145/1345206.1345215.
[64] J. Wang, K. Zhang, X. Tang, and B. Hua, "B-Queue: Efficient and Practical Queuing for Fast Core-to-Core Communication," *Internation Journal of Parallel Programming*, vol. 41, pp. 137-159, 2023, DOI: 10.1007/s10766-012-0213-x.
[65] Boost.Lockfree, 2024, [Online]. Available: https://www.boost.org/doc/libs/1_85_0/doc/html/lockfree.html.
[66] R. Marotta *et al.*, "Mutable locks: Combining the best of spin and sleep locks," *Concurrency and Computation: Practice and Experience*, vol. 32, no. 22, 2020, DOI: 10.1002/cpe.5858.